\documentstyle[pre,aps,psfig]{revtex}

\def\be{\begin{equation}} 
\def\ee{\end{equation}}
\def\pa{\partial}
\def\na{\nabla}

\def\rarrow{\rightarrow}

\draft
\begin{document}

\title{Some properties of a class of coupled Langevin equations }

\author{Parthapratim Biswas\footnote{ppb@boson.bose.res.in},
Arnab Majumdar\footnote{arnab@boson.bose.res.in}, 
Anita Mehta\footnote{anita@boson.bose.res.in}}

\address{ S.N.Bose National Centre For Basic Sciences\\
Block JD, Sector III,
Salt Lake City, Calcutta-700 091, INDIA}

\author{J.K.Bhattacharjee\footnote{tpjkb@iacs.ernet.in}}

\address{ Department of Theoretical Physics\\
Indian Association For Cultivation of Sciences\\
Jadavpur,Calcutta-700 032, INDIA}

\maketitle

\begin{abstract}

We present a set of coupled continuum equations
with a specific coupling between mobile grains $\rho$
and clusters $h$  on the surface of a sandpile. The
equations are analysed self-consistently; we demonstrate that 
Edwards' infrared divergence is responsible for the unexpected critical exponents
we find, which are verified by simulations.

\end{abstract}

A certain class of coupled Langevin equations \cite{kn:ambook,kn:mln,kn:bcre}
has been useful of late in describing the dynamics of sandpile surfaces 
\cite{kn:rpp}. 
The fluctuations on the
sandpile surface are described by the local height $h(x,t)$ and the density
$\rho(x,t)$ of flowing grains -- the flow being initiated by the formation of
bumps. The coupling between the two variables is described in terms of a 
transfer term which converts static grains to mobile grains and vice-versa.
In this communication we point out that a singularity 
discovered by Edwards \cite{kn:edw}
three decades ago in the context of fluid turbulence is present in such 
models due to the specific choice of the transfer term. This singularity 
largely controls the dynamics and produces unexpected exponents. Our 
contention is supported by numerical evidence.

We begin with a model introduced by Mehta, Luck and Needs (MLN) \cite{kn:mln} 
where the coupled Langevin equations are  
\begin{eqnarray}
\label{eq-cup}
{\pa h \over \pa t}    &=& D_h\na^{2} h -T(h,\rho) + \eta_{h}(x,t) \nonumber \\
{\pa \rho \over \pa t} &=& D_{\rho}\na^{2} \rho +T(h,\rho) +\eta_{\rho}(x,t) \\
T(h,\rho) &=& -\mu\rho({\na h)} \nonumber
\end{eqnarray}
where the terms $\eta_h(x,t) $ and $\eta_{\rho}(x,t) $
represent  
Gaussian white noise.
A variant of the above is the model due to Bouchaud {\it et al.} (BCRE)
\cite{kn:bcre} where
\[ T = -\nu \na h  - \mu \rho (\na h) \]
and the noise is present only in the equation of motion for $h$. All
the results which we will find here for the MLN system   
also holds for BCRE which we have checked both analytically and numerically.

A simple physical picture of the coupling or `transfer' term 
$T(h,\rho)$ between
$h$ and $\rho$ is the following: flowing grains
are added to regions of the interface
which are at less  than the critical slope, and vice versa,
{\it provided that the local density of flowing grains
is always non-zero} and in proportion to the local density.
This is the simplest possible form of
exchange between the species that appears
intuitively reasonable during the process of
avalanching and we analyse  in what follows the
profiles of both species consequent on this form.

Before this, we review some well-known
facts about interfacial roughening \cite{kn:interf}.
Three critical exponents, $\alpha$, $\beta$, and $z$,
characterise the spatial and temporal scaling
behaviour of a rough interface.
They are conveniently defined by considering the (connected)
two-point correlation function of the heights
\[
S(x- x^{\prime},t-t^{\prime})=\bigl\langle h(x,t)h(x^{\prime}
,t^{\prime})\bigr\rangle
-\bigl\langle h(x,t)\bigr\rangle\bigl\langle h(x^{\prime},t^{\prime})\bigr\rangle. \nonumber
\]
We have
\[
S(x,0)\sim \vert x\vert^{2\alpha}\quad(\vert x\vert\to\infty)
\quad  \mbox{and} \quad S(x=0,t)\sim \vert t\vert^{2\beta}\quad(\vert t\vert\to\infty),\nonumber
\]
and $z = \alpha/\beta$.

 In recent years a self-consistent mode
coupling analysis used hitherto in dynamic critical phenomena  \cite{kn:kawasaki}
has been used to look at in particular the Kardar-Parisi-Zhang (KPZ) 
equation \cite{kn:kpz}
and we extend its use to the case of the coupled equations presented here. 

In this method we set up equations (to one-loop order) for the correlation 
functions and self-energies in terms of the full Green's functions,
 correlation functions
and vertices using assumed scaling forms for each. The critical exponents 
$\alpha$ and $\beta$ defined above are obtained from
the self-consistent solutions of these equations after setting $D_h$ and 
$D_\rho$ to unity.

  The analysis of these functions will be in terms of a weak scaling hypothesis
which states 
\begin{eqnarray}
  G_h(k,\omega) = k^{-z_h} f_h\left(\frac{\omega}{k^{z_h}},\frac{\omega}{k^{z_\rho}}\right)\nonumber
\end{eqnarray}
along with a similar scaling relation for $G_\rho(k,\omega)$.
 A strong scaling would imply the existence of a single time scale $i.e.$ $z_h
= z_\rho$. As we show below, this cannot be the case here. The absence of strong 
scaling implies that the roughness exponents $\alpha_h$ and $\alpha_\rho$ may 
become functions of $k$.

We consider the full Green's function $G_h(k,\omega)$, which is
given via the well-known Dyson equation,
\[
G_h^{-1}(k,\omega) = {G_h^{0}}^{-1}(k,\omega) + \Sigma_h(k,\omega)
\] 
Here, the zeroth order Green's function is 
\[
G_h^{0}(k,\omega) = (-i\omega+k^2)^{-1} \nonumber
\]

In the scaling limit $k^2$ can be dropped in comparison with $\Sigma_h(k,\omega)$.
To one-loop order, the self-energy $ \Sigma_h(k)$ is given by 
\begin{eqnarray}
\Sigma_h(k,\omega) & = & \mu^{2} (2\pi)^{-2} \int dq \int d\Omega
G_h(k-q,\omega-\Omega)S_{\rho}(q,\Omega) \, k(k-q) \label{eq-seh1} 
\end{eqnarray} 
We note that due to the presence of the term $S_\rho(q,\Omega)$, the
integral is dominated by the 
singularity in the integrand at $q \rightarrow 0 $. This `infrared divergence'
which relates to the divergence of the {\it internal} momenta $q$, 
is very different from the usual divergences encountered
in critical phenomena where the latter occur 
for small wave numbers and are associated with
long wavelength instabilities in the external momenta. In this case due to  
the infrared divergence in the above equation in the internal momenta $q$,
the integral diverges {\it for any value of the external momenta $k$},
 so long as $\alpha_{\rho}>0$. This is the divergence found by Edwards for
the Navier-Stokes equation \cite{kn:edw}.
We thus need either to evaluate the integral with a lower cut-off $k_0$ or to
introduce a suitable regulator. We follow the first of these procedures
for the above equation.

We then proceed to evaluate the self-energy at zero external frequency, $i.e$
$\Sigma_h ( k, \omega=0) $ from Eq.(\ref{eq-seh1}).
The integral in Eq.(\ref{eq-seh1}) becomes in the limit of zero external 
frequencies  
$$ \Sigma_h(k) = {\mu^{2}k^{2}\over \Sigma_h(k)} \int {dq \over 2\pi}
\int {d\Omega \over 2\pi} S_{\rho}( q,\Omega)  $$
We have to evaluate the integral by cutting off the momentum integration
at $k_{0} \ll 1$ , $i.e.$ we follow the first of the procedures
given above to handle the infrared divergence. This gives, after
some simplification,
\[
{\Sigma_h^{2}(k)}= \mu^{2}k^{2} {k_{0}^{-2\alpha_{\rho}}
C_{\rho}\over 4\pi\alpha_\rho} \nonumber
\]
From the above equation with the scaling relation $\Sigma_h(k) \sim k^{z_h} $
we find, on equating powers of $k$,
$$ z_h = 1$$
We note here that the presence 
of the term $\rho\na h$ 
could in principle cause the
vertex $\mu$ to renormalize,
leading to a correction to 
$ z_h$. We have checked that this correction vanishes as $k \rarrow 0$ in
the lowest order.

The structure factor at one-loop level is given by
\begin{eqnarray}
S_h(k,\omega) & = & \frac{1}{\omega^2 + |\Sigma_h(k,\omega)|^2}
 \left[ 1 + \mu^2\int{dq\over 2\pi}\int{d\Omega\over 2\pi}  
 |k-q|^2 S_h(k-q,\omega-\Omega) S_\rho(q,\Omega)\right]
\label{eq-shkw}
\end{eqnarray}
which 
on integrating with respect to $\omega$ gives $S_h(k,t=0)$ as
\be S_h(k,t=0) \equiv \int S_h(k,\omega) {d\omega \over 2\pi} = \frac{A_0}{k} + \frac{B_0}{k^{1+
2\alpha_h}} \label{eq-shkt} \ee
Recognising that  the scaling form of $S_h(k,t=0) \sim k^{-1-2\alpha_h}$, we
notice that $\alpha_h$ cannot in general be determined  from  Eq.(\ref{eq-shkt}).
This is because the second term on the right-hand-side of Eq.(\ref{eq-shkt})
dominates at small momenta $k$ provided $\alpha_h > 0 $.  

We turn now to the critical exponents
in $\rho$. The single loop 
self-energy $\Sigma_{\rho}(k,\omega)$ is given by
\be
\Sigma_{\rho}(k,\omega=0) = -\mu^{2}(2\pi)^{-2}\int dq
\int d\Omega G_{\rho}(k-q,-\Omega)S_h(q,\Omega) q^{2} 
\ee
This gives, on performing the integral over internal frequency $\Omega$,
\be
\Sigma_{\rho}(k,\omega=0) = -\mu^{2}\int {dq \over 2\pi}{q^2 \over q^{1+2\alpha_h}}
{1\over {\vert k-q \vert}^{z_\rho} + q^{z_h}} \nonumber
\ee
We see from the above that
$\Sigma_\rho(k,0)$, the relaxation rate for $\rho$ fluctuations, is 
negative and finite as $k \rightarrow 0$, 
 and we  need to add a positive constant, $\Sigma_0$, to the
self-energy ($\Sigma_0 > |\Sigma_\rho(k \rightarrow 0)|$) for regulatory purposes. 
This divergence in the relaxation rate, needing
regulation, is reflected in the divergence we have encountered
in our numerical investigations below;  we have there
followed an analogous procedure by introducing a numerical
regulator which replaces divergent values of the transfer
term by suitably defined cutoffs \cite{kn:mln}.
The resulting constancy of $\Sigma_{\rho}$ implies $z_{\rho} \approx 0 $ 
for the regulated equations and will be used in the following.
  
The correlation function $S_{\rho}(k,\omega)$ is given by 
\begin{eqnarray}
S_{\rho}(k,\omega) & = & (\omega^{2}+k^{2z_\rho})^{-1} (2\pi)^{-2}\int dq 
\int d\Omega(k-q)^2 S_h(k-q,\omega-\Omega)S_\rho(q,\Omega) 
\end{eqnarray} 
which on itegration over $\omega$ gives
$$ S_\rho(k,t=0) \sim k^{-(1+2\alpha_\rho)}  \sim  
k^{1-2\alpha_h+z_h}{1\over k^{z_\rho + z_h } ( k^{z_h}+k^{z_\rho} )}$$
Finally using $z_{\rho} \approx 0 $ we have 
\begin{eqnarray}
\alpha_{\rho} = \alpha_h+ {z_h\over 2}-1 & \quad & \mbox{for large $k$}\label{eq-rho1} \\
\alpha_{\rho} = \alpha_h-1  & \quad & \mbox{for small $k$}\label{eq-rho2} 
\end{eqnarray}

Given our numerical result of $\alpha_h = 0.5$,
the above predicts a negative $\alpha_\rho$,
at small $k$.
This is consistent with, and validates
our assumption of, a cutoff $k_0$
which arises naturally as the wavevector
separating the region of $\alpha_\rho < 0$
and
$\alpha_\rho > 0$.

The coupled equations have been
numerically integrated by using the method of finite differences. 
Our grids in time and space were kept as fine-grained as
computational constraints allowed. This is in order to avoid the
instabilities associated
with the discretisation of nonlinear continuum equations.
 Convergence has been checked 
by keeping $\Delta t $ small enough such that the quantities under
investigation are independent
of further discretisation. In all our calculations, we chose $ D_h =
D_{\rho} = \mu = 1$.

On discretising the equations Eqs.(\ref{eq-cup})
we found once again the divergences
that were previously observed in \cite{kn:mln}.
These divergences are in our
view a direct representation
of the negativity of $\Sigma_\rho$. We follow
here a parallel course in regulating
these via an explicit regulator.
In earlier work \cite{kn:mln}, a regulator
was introduced which replaced
the function $\mu\rho\na h$ 
by the following:
\begin{eqnarray}
  T &=&  +1  \quad\quad\quad\mbox{for}\quad \mu \rho (\na h) > 1 \nonumber \\
    &=&  \mu \rho (\na h)   \quad \mbox{for} \quad -1\le \mu\rho(\na h) \le 1 \nonumber\\
    &=&  -1  \quad \quad\quad\mbox{for} \quad \mu \rho (\na h) < -1  \nonumber
\end{eqnarray}
In addition in this paper,
we have introduced noise reduction
to the regulated equations
which has led to a more accurate
evaluation of all our critical
exponents.

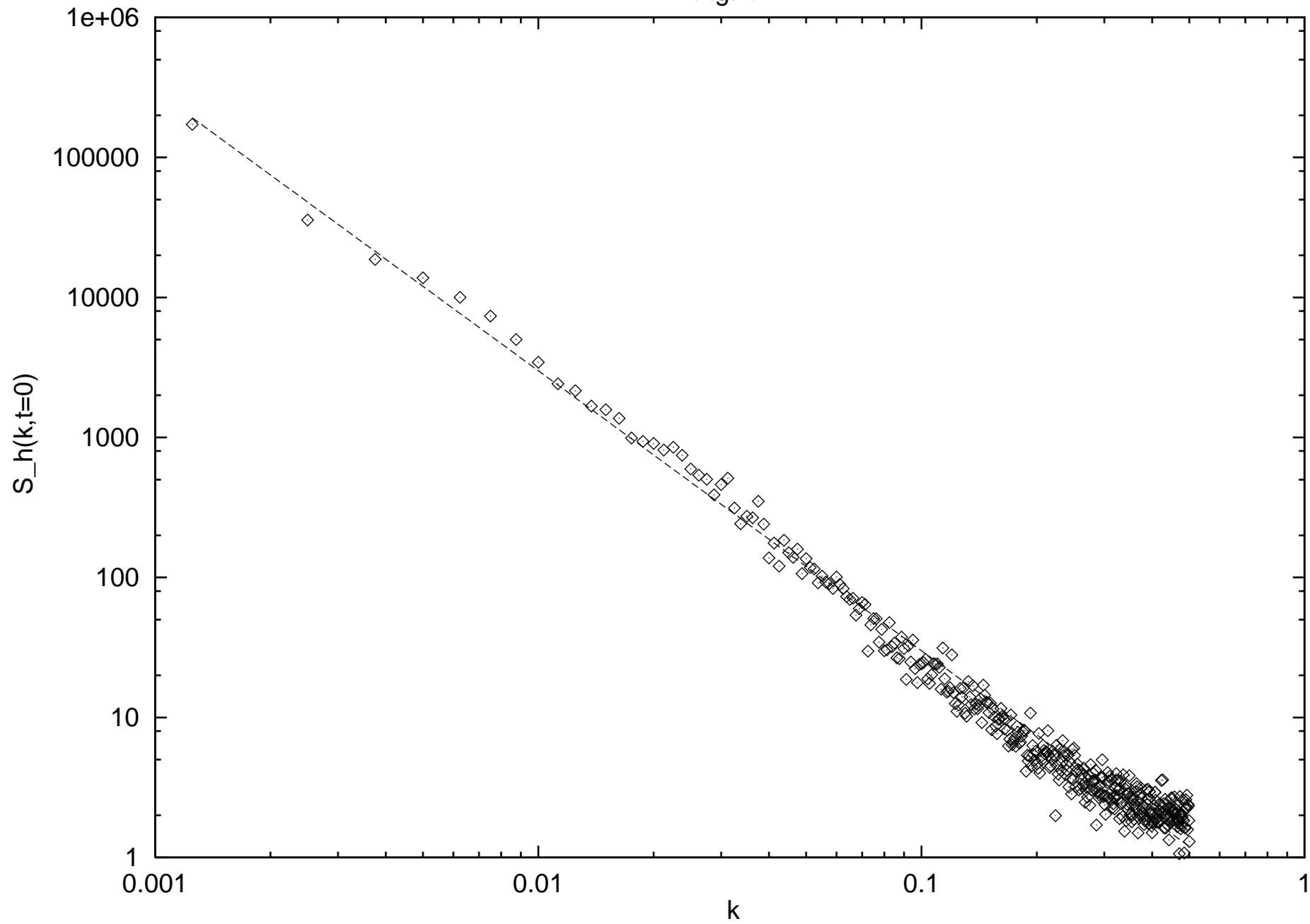
\begin{figure}
\caption{
The single Fourier transform $S_h(k,t=0)$. 
with a power-law fit to an exponent of $2.0\pm0.04$.}
\label{fig-1}
\end{figure}

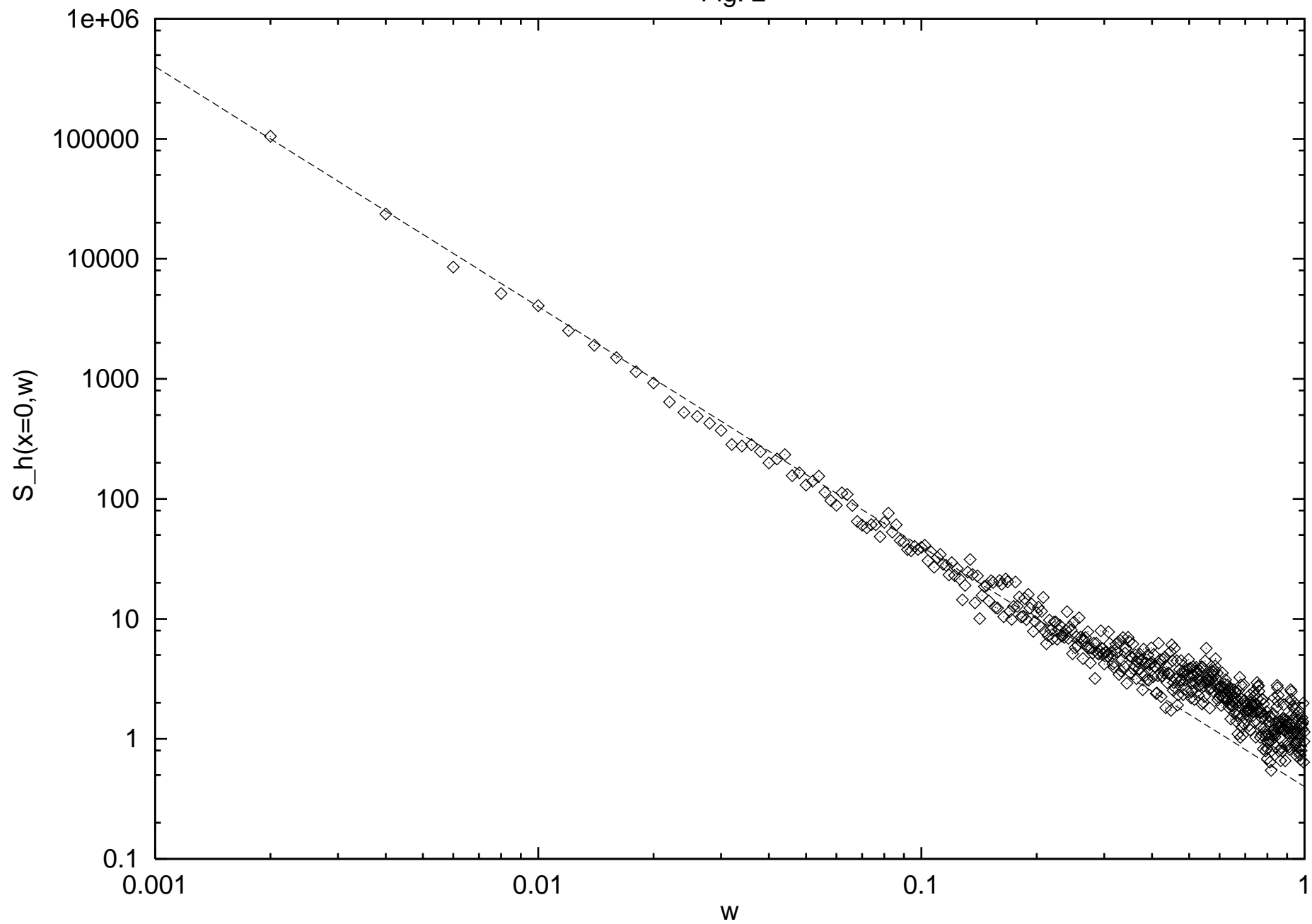
\begin{figure}
\caption{
The single Fourier transform $S_h(x=0,\omega)$. 
with a power-law fit to an exponent of $1.96\pm0.02$.}
\label{fig-2}
\end{figure}

Our results for the single Fourier 
transforms for the $\langle hh \rangle$ correlation function
are \\
(i) 
The Fourier transform 
$ S_h(k,t=0) $ (Fig.\ref{fig-1}) 
 is consistent with
a spatial roughening exponent $\alpha_h \sim 0.5\pm 0.02   $  
via  our observation of
$ S_h(k,t=0) \sim {k^{-2.00\pm 0.04 }} $\\
(ii)
The Fourier transform 
$ S_h(x=0,\omega) $ (Fig.\ref{fig-2}) 
 is consistent with
a temporal roughening exponent 
$\beta \sim 0.48\pm 0.01 $
via  our observation of
$ S_h(x=0,\omega) \sim {\omega^{-1.96\pm 0.02 }} $
Hence
$ z_{h}\sim 1.00\pm 0.02$ consistent with our prediction.

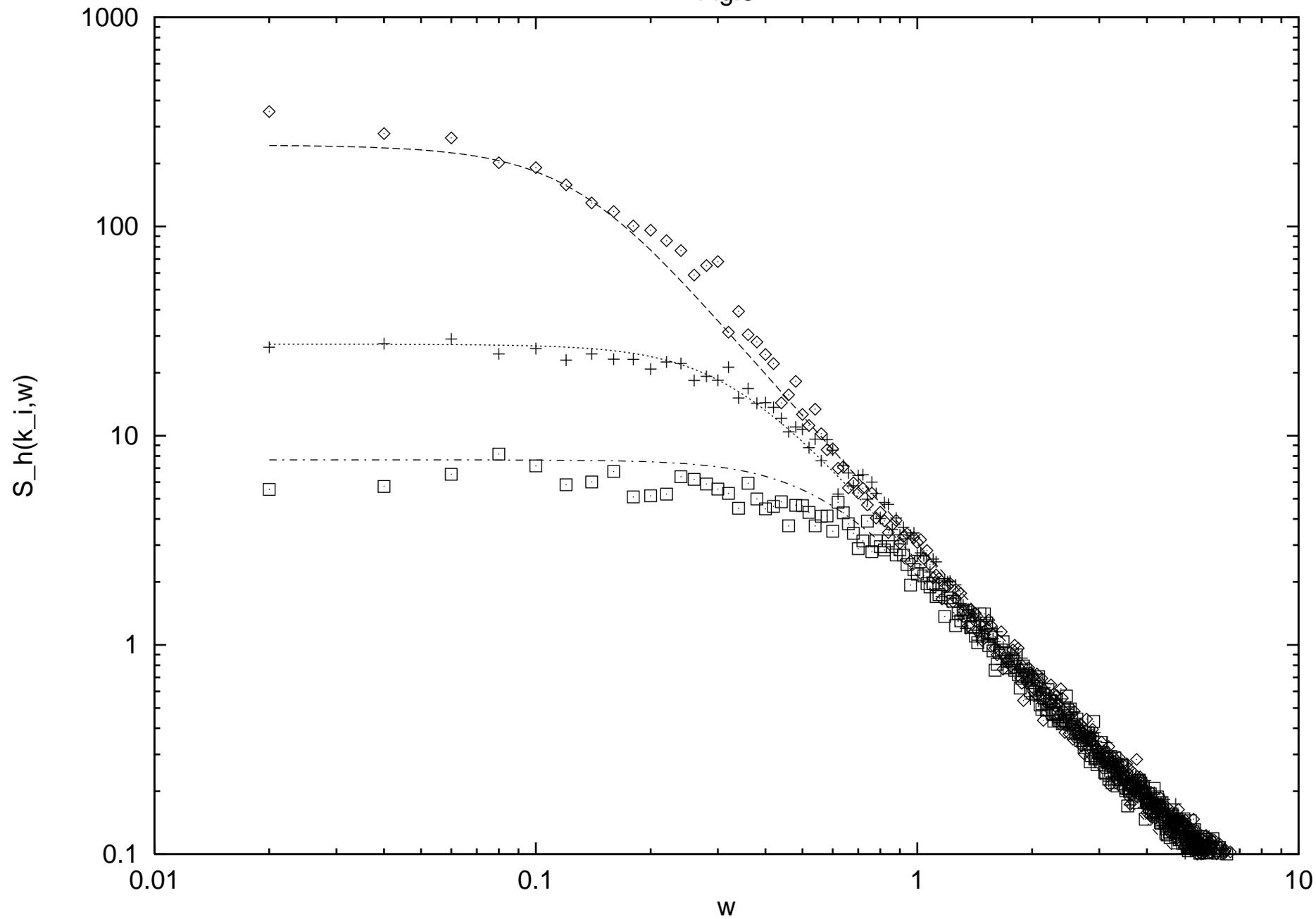
\begin{figure}
\caption{
The double Fourier transformrs $S_h(k_i,\omega)$ evaluated at 
three different $k_i = 0.1, 0.2, 0.3$ 
fitted to Eq.(\ref{eq-shkw}).}
\label{fig-3}
\end{figure}

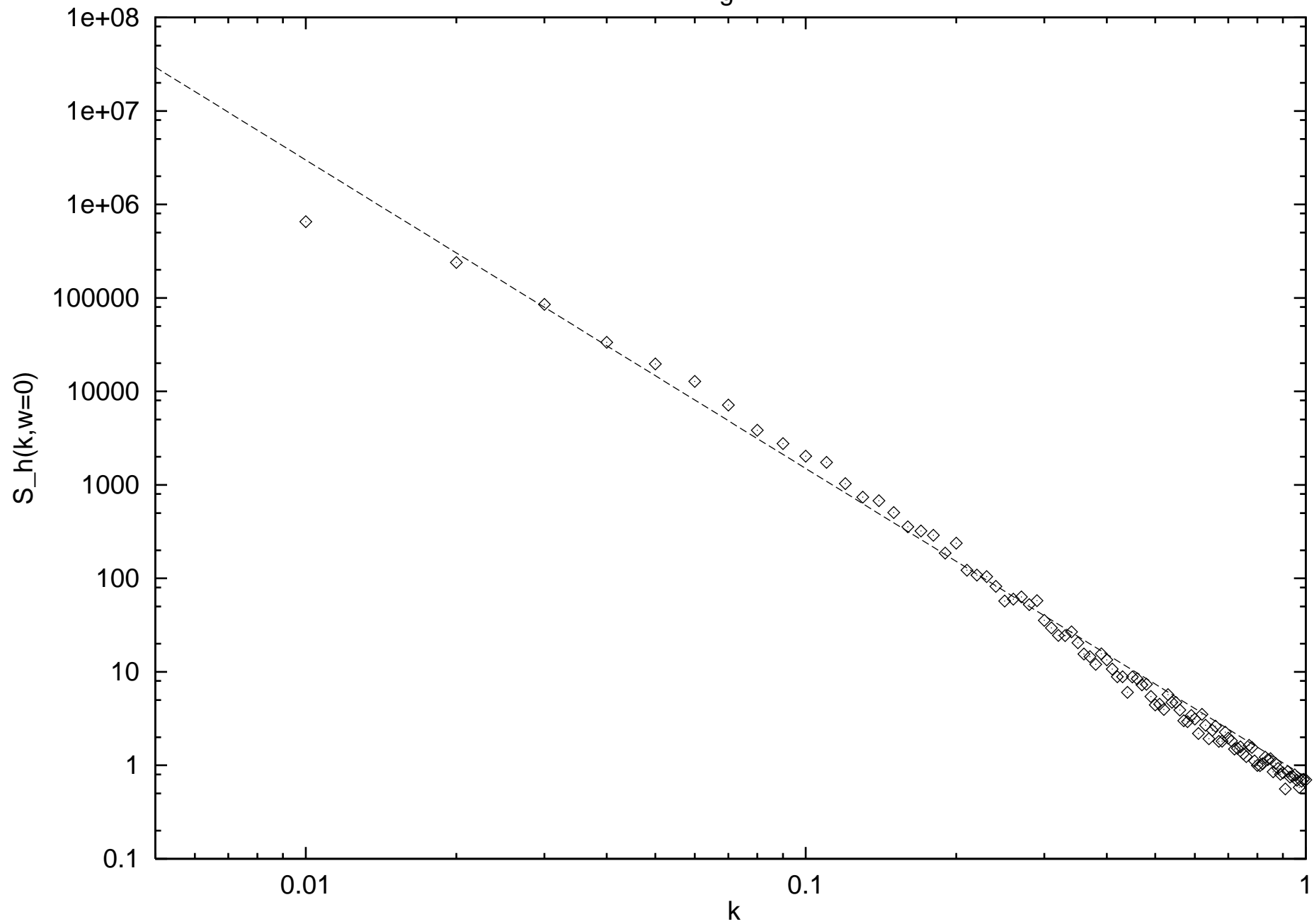
\begin{figure}
\caption{
The double Fourier transform $S_h(k,\omega=0)$ with a power-law fit to
an exponent of $3.3\pm0.05$.}
\label{fig-4}
\end{figure}

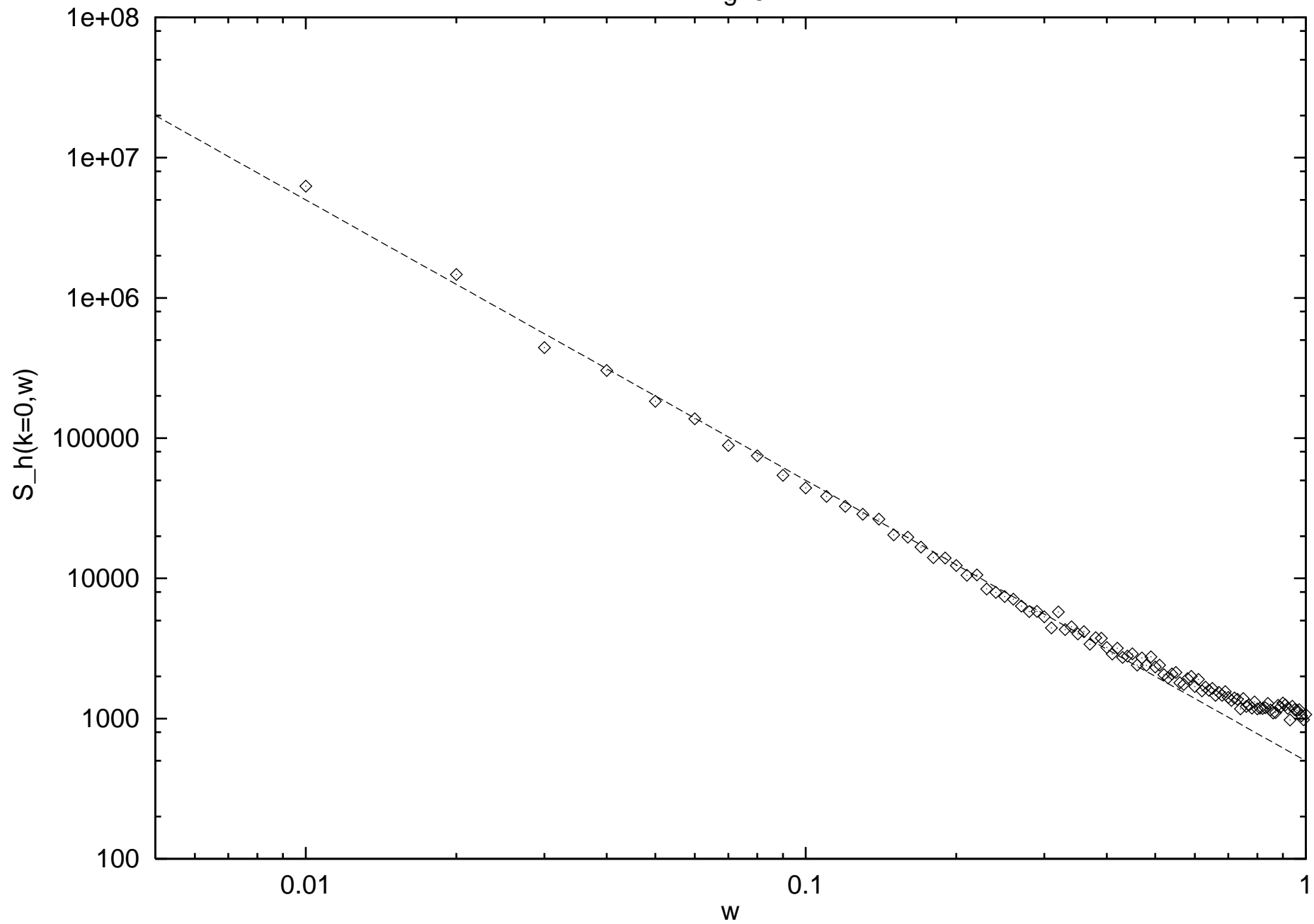
\begin{figure}
\caption{
The double Fourier transform $S_h(k=0,\omega)$ with a power-law fit to
an exponent of $1.97\pm 0.03$.} 
\label{fig-5}
\end{figure}

The full structure factor $S_h(k,\omega) $ has been 
calculated at three different $k $ points and  Fig.\ref{fig-3}   
displays a fit of our results to an appropriately scaled form of Eq.(\ref{eq-shkw}). 
The spatial structure factor
 $ S_h(k,\omega=0)$ shows a power-law
behaviour (Fig.\ref{fig-4}) given by 
 $ S_h(k,\omega=0) \sim {k^{-3.30 \pm .05}} $
in qualitative accord with our expression, which predicts an exponent of $-3$.
The temporal structure factor
 $ S_h(k=0,\omega)$ shows a power-law
behaviour (Fig.\ref{fig-5}) given by 
$ S_h(k=0,\omega) \sim {\omega^{-1.97\pm .03}} $
which is in agreement with our prediction of $\omega^{-2}$ from the expression of the 
structure factor (Eq.(\ref{eq-shkw})).

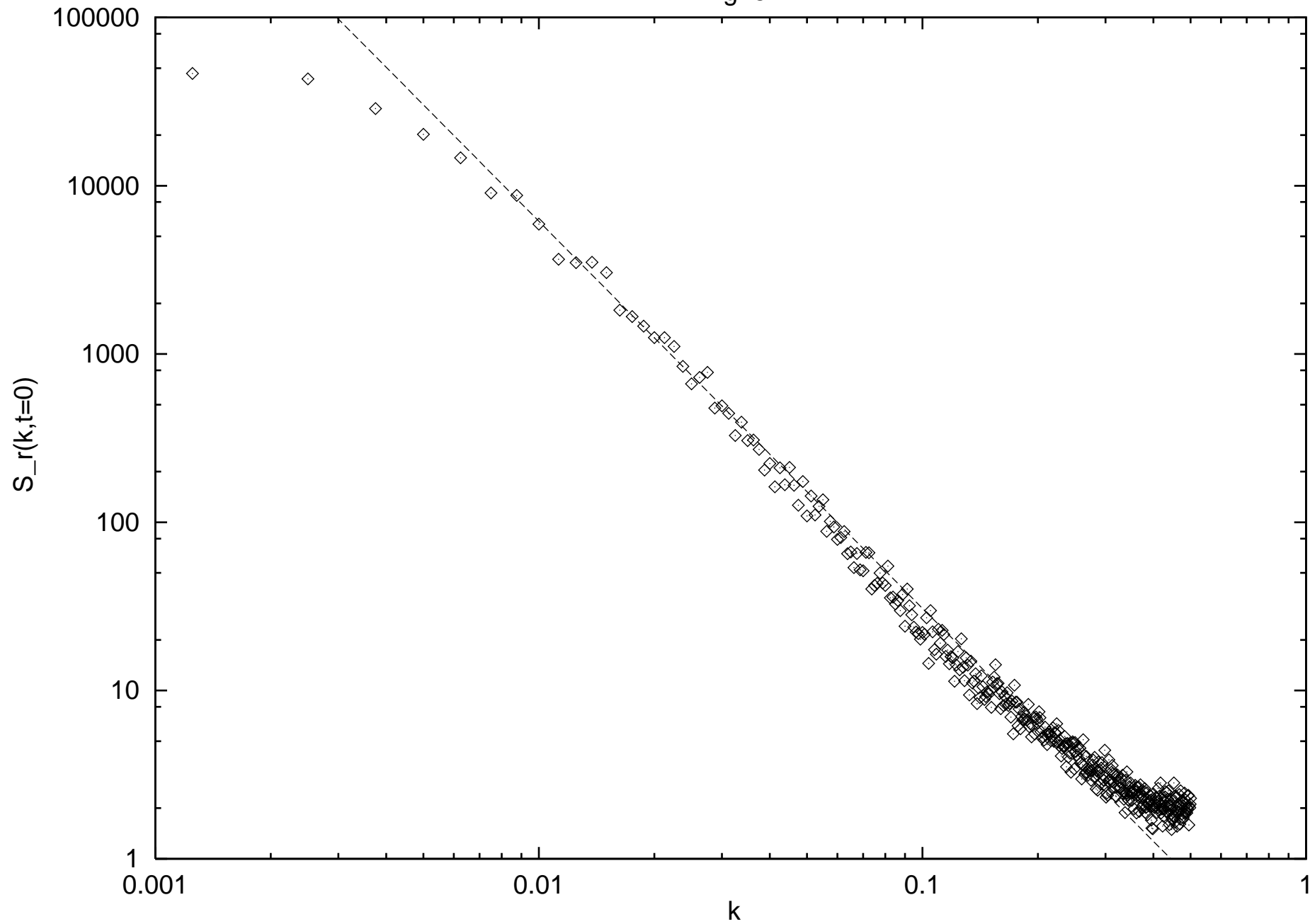
\begin{figure}
\caption{ 
The single Fourier transform $S_\rho(k,t=0)$ 
with a power-law fit to an exponent of $2.0 \pm 0.08$ at large $k$
crossing over to an exponent of zero at small $k$.}
\label{fig-7}
\end{figure}

Given our values of $\alpha_h \simeq 0.5$ and $z_h \simeq 1$,
Eqs.(\ref{eq-rho1}) and (\ref{eq-rho2}) 
predict a crossover in $\alpha_\rho$ from 0.0 at large $k$ to
-0.5 as $k \rarrow 0$.
We observe that the single Fourier transform 
 $S_\rho(k,t=0)$
(Fig.\ref{fig-7})  
 shows a crossover behaviour from 
$ S_\rho(k,t=0)\sim {k^{-2.00\pm 0.08}} $ 
for large wavevectors to
$ S_\rho(k,t=0)\sim \mbox{constant} $ 
as $k \rarrow 0$. 
 This indicates a crossover from $\alpha_\rho = 0.5$ for large $k$
to -0.5 as $k \rarrow 0$, which shows the same trend as the prediction above.
Note however that the simulations
also manifest in addition to the above 
the normal diffusive behaviour represented
by $\alpha_\rho =0.5 $ at large wavevectors.
This anomalous smoothing behaviour in $\rho$ is the direct consequence of the infrared 
divergence in Eq.(\ref{eq-seh1}) discussed in the preceding.

We have analysed in the above via perturbative methods a model of
sandpile dynamics which was presented without analysis
in earlier work \cite{kn:mln}.  The good agreement between our
theoretical predictions and numerical simulations confirms our
contention that Edwards' infrared divergence \cite{kn:edw} plays a 
crucial role in producing unexpected critical exponents in our model.

\end{document}